\documentclass[prb, twocolumn]{revtex4-2}

\usepackage[english]{babel}
\usepackage[utf8]{inputenc}
\usepackage{graphicx}
\usepackage[intlimits, sumlimits, namelimits]{mathtools}
\usepackage{physics}
\usepackage{mathrsfs}
\usepackage{xcolor}

\begin{document}
\title{Interlaced wire medium with quasicrystal lattice}

\author{Eugene~A.~Koreshin${}^{1}$}
\email{evgeniy.koreshin@metalab.ifmo.ru}
\author{Mikhail~V.~Rybin${}^{1,2}$}

\affiliation{$^1$School of Physics and Engineering, ITMO University, St Petersburg 197101, Russia}
\affiliation{$^2$Ioffe Institute, St Petersburg 194021, Russia}

\date{\today}

\begin{abstract}
We propose a design of interlaced wire medium with quasicrystalline lattice based on five-fold rotation symmetry Penrose tiling. The transport properties of this structure are studied. We distinguish two transport regimes, namely, propagation regime related to the low-frequency interval and localization regime in the high-frequency interval. While the former is observed in structures both with and without translation symmetry property, the latter is exclusive for aperiodic structures only. We show that the localization regime is promising for many applications including engineering of effective multi-channel devices for telecommunication and imaging systems.
\end{abstract}

\maketitle

\section{Introduction}

Interlaced wire media (IWM) \cite{shin2007three} have been in the focus of attention in the past decade due to unique properties of wave transport \cite{latioui2017light,sakhno2021longitudinal} because of the rapid phase variation along the propagation direction in a structure consists of more than two independent interpenetrating metallic grids. Many reports have considered, among other phenomena, the broadband negative group velocity in the low-frequency range with coupling effect depending on propagation direction \cite{chen2018metamaterials}, the longitudinal nature of electromagnetic waves supported by double grids IWM \cite{latioui2017light}, the rotation of polarization and phase control \cite{shin2010transmission,powell2021dark}, the design of graded refractive index lenses \cite{wang2021three} and IWM with an arbitrary large number of photonic bands with a linear dispersion relation in the low-frequency limit \cite{shin2007three}.

The most convenient way for analysis of the structure is to exploit translation symmetry leading to Bloch solutions.  This allows using many powerful concepts of solid-state physics such as reciprocal space, wave vectors, Brillouin zone. However, translation symmetry inherently limits the possible transport properties attributed to waves with either purely real or complex wavenumbers. In particular, no intrinsic localization of waves occurs in periodic systems. Disordered structures enable a number of transport phenomena, but they only make sense for values under statistical averaging over a large ensemble of various occurrences.  

Quasicrystals are between the two extreme cases. They have a rigorous ordering of their elements without any translation symmetry. Recent research in the field of electrodynamics of quasicrystals \cite{vardeny2013optics} has revealed some interesting phenomena that are not related to periodicity including Dirac conical dispersion \cite{dong2015conical}, intrinsic wave localization \cite{jeon2017intrinsic, sinelnik2020experimental}, quantum spin Hall effect \cite{huang2018quantum,huang2018theory,huang2019comparison}, superconductivity \cite{araujo2019conventional}, existence of surface plasmon \cite{yang2016near}, phenomena in topological Anderson \cite{chen2019topological} and Chern \cite{he2019quasicrystalline,duncan2020topological} insulators. Noteworthily, the absence of the condition for exploiting the Bloch theorem results in the quasicrystal analysis being based in most cases on simulation of the wave equations over large samples usually of the size of dozens wavelengths.

\begin{figure}[b]
\includegraphics[width=\columnwidth]{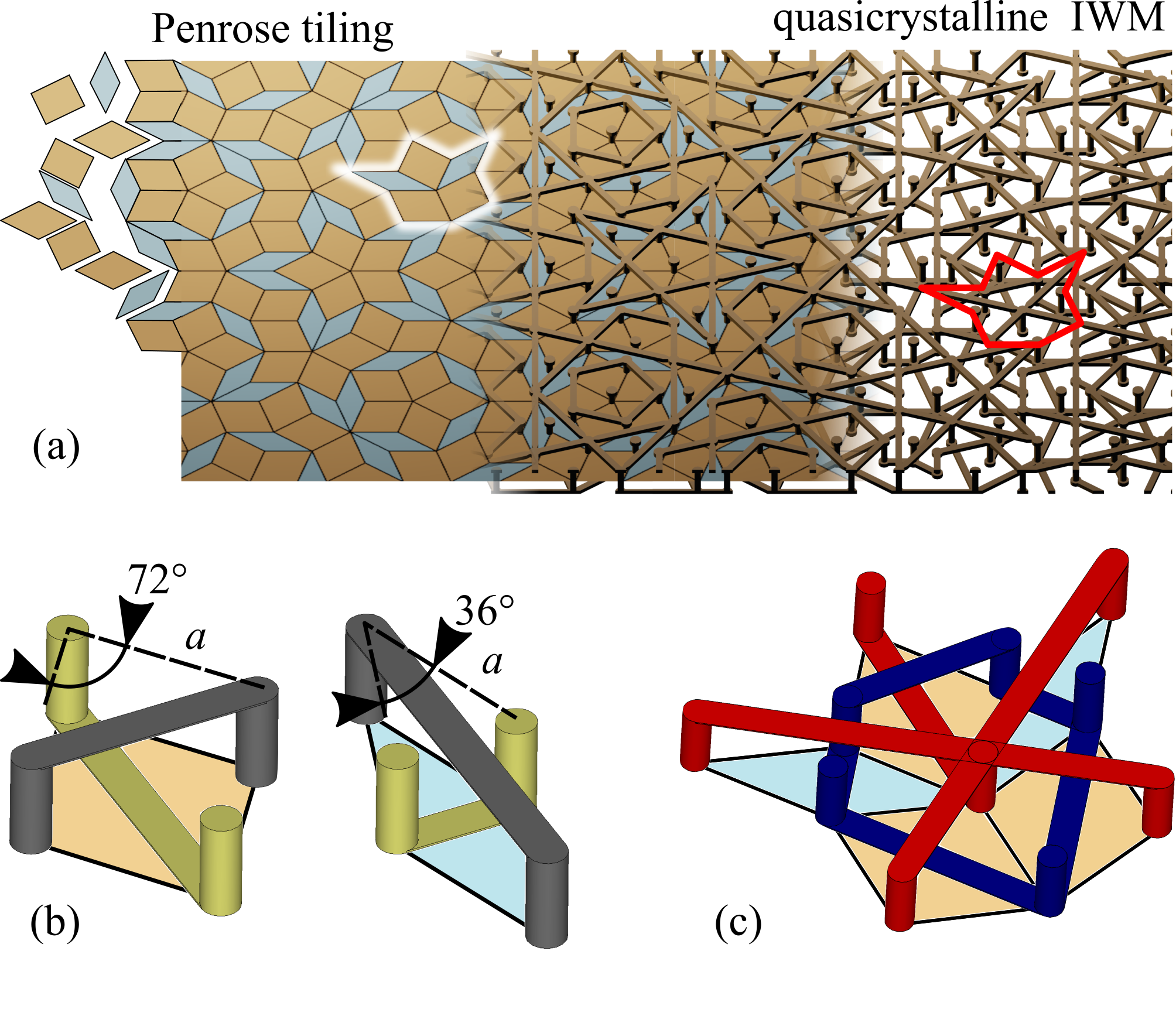}
\caption{\label{fig:geometry}
Design of quasicrystalline interlaced wire medium based on Penrose tiling pattern (a).
(b) Two unit cells of Penrose tiling that are thick and thin parallelepipeds (based on rhombuses), which include two non-connected wires disposed in the top (gray) and bottom (yellow) planes.
(c) Five tiles attached to the same node. The first subgrid (shown in red color) includes the wires joined to the central node, the second one (blue) is formed by the wires surrounding the central node. The types of tiles are shown in the bottom plane. 
The star (c) is highlighted twice on (a).
}
\end{figure}
\begin{figure*}
		\includegraphics[width=\linewidth]{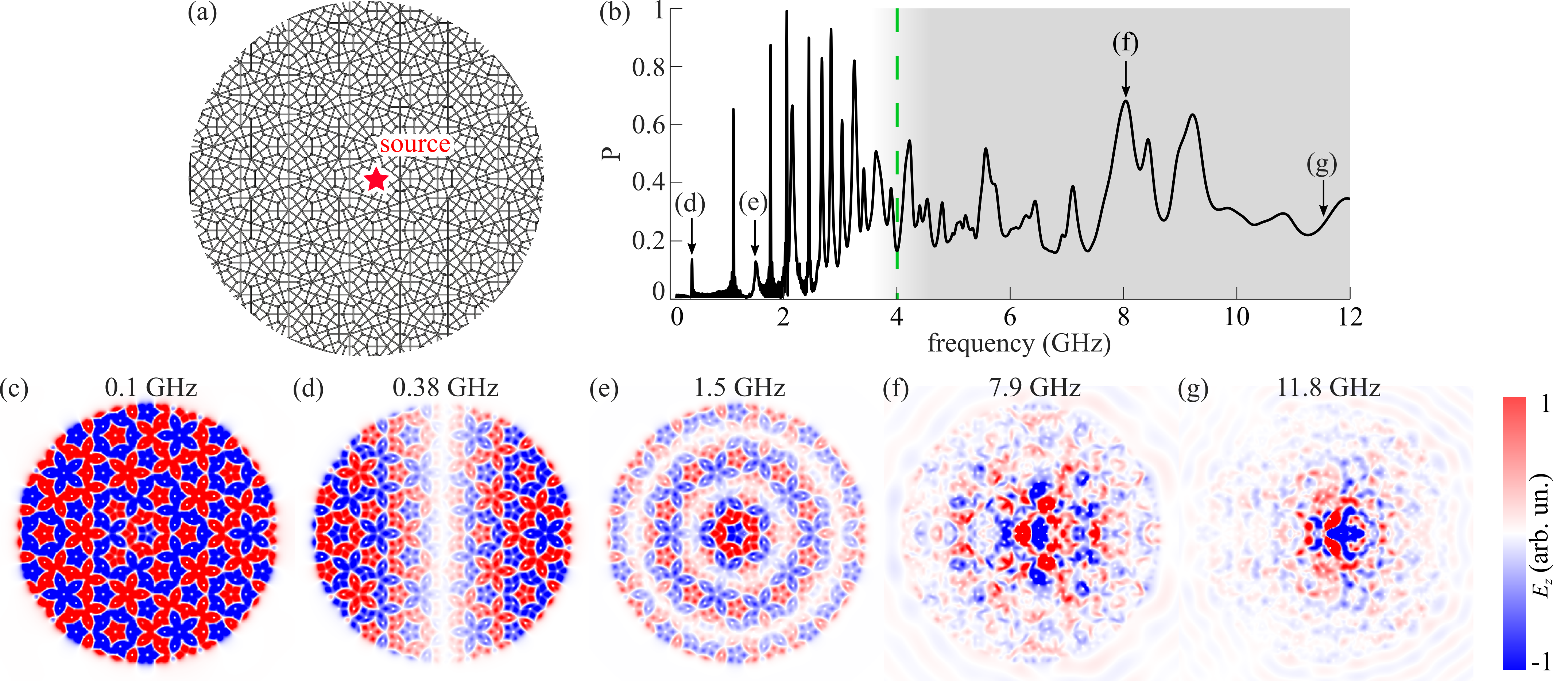}
			\caption{\label{fig:modes} 
			(a) The finite sample of the quasicrystal consisting of 665 tiles. The star marks the source connected to two grids. The source has vertical orientation for generating $z$-polarisation of the electric field. (b) The spectrum of power losses. The transport regimes of propagation (low-frequency interval) and localisation (high-frequency interval) are separated by the shadings. Electric field $E_z$ distribution for quasistatic (c), propagation (d-e) and localization (f-g) regimes.}
\end{figure*}

The metallic wire grids forming IWM are supposed to have a complex topology in three-dimensional (3D) space for the separate grids not to be connected to each other. Although weak deviation from the periodic arrangement of IWM is easily implemented, observation of aforementioned effects related to the absence of translation symmetry requires either a very large sample, the size of which is limited by restricted computing facilities or a strong decline from the translation symmetry typical for quasicrystals. The latter is a challenging problem since it involves both restrictions of IWM topology and quasicrystal spatial ordering which have to be met simultaneously.

In this paper, we propose a design of IWM in an aperiodic structure based on Penrose-type quasicrystal with the $C_5$ rotation axis symmetry. We study wave transport properties of the designed IWM system by using full-wave simulation of the electromagnetic problem. We discover three transport regimes, namely, the  static and propagation that are similar to IWM with translation symmetry and also an exclusive quasicrystal regime of wave localization. The latter provides the a benefit of multi-channel antenna design with large decoupling coefficients.

\section{Design of quasicrystalline interlaced wire medium}

Let us recall the conventional Penrose tiling of two-dimensional space. The quasicrystal is composed of two different tiles (or unit cells) that are a pair of rhombuses with the same edge length. The thick one has the angles of 72$^\circ$ and 108$^\circ$ and the thin one has the angles of 36$^\circ$ and 144$^\circ$ (see Fig.~\ref{fig:geometry}b). Regular quasicrystal structure is supposed to cover the entire plane by closely packing the two unit cells with no gaps between them. Thus, each node has to be attached to the rhombus, the apex angles of which the sum of 360$^\circ$. This requirement leads to 54 possible stars (a cycle combination of the rhombuses around their common node). Nevertheless, due to the matching rules \cite{grunbaum1987tilings}, only 7 combinations remain. Several approaches have been developed to fit the plane with Penrose tiling by two unit cells \cite{janot1996quasicrystals}. In order to generate a large-size tiling here we exploit the up-down approach \cite{senechal1996quasicrystals}. 

\begin{figure*}
\includegraphics[width=\linewidth]{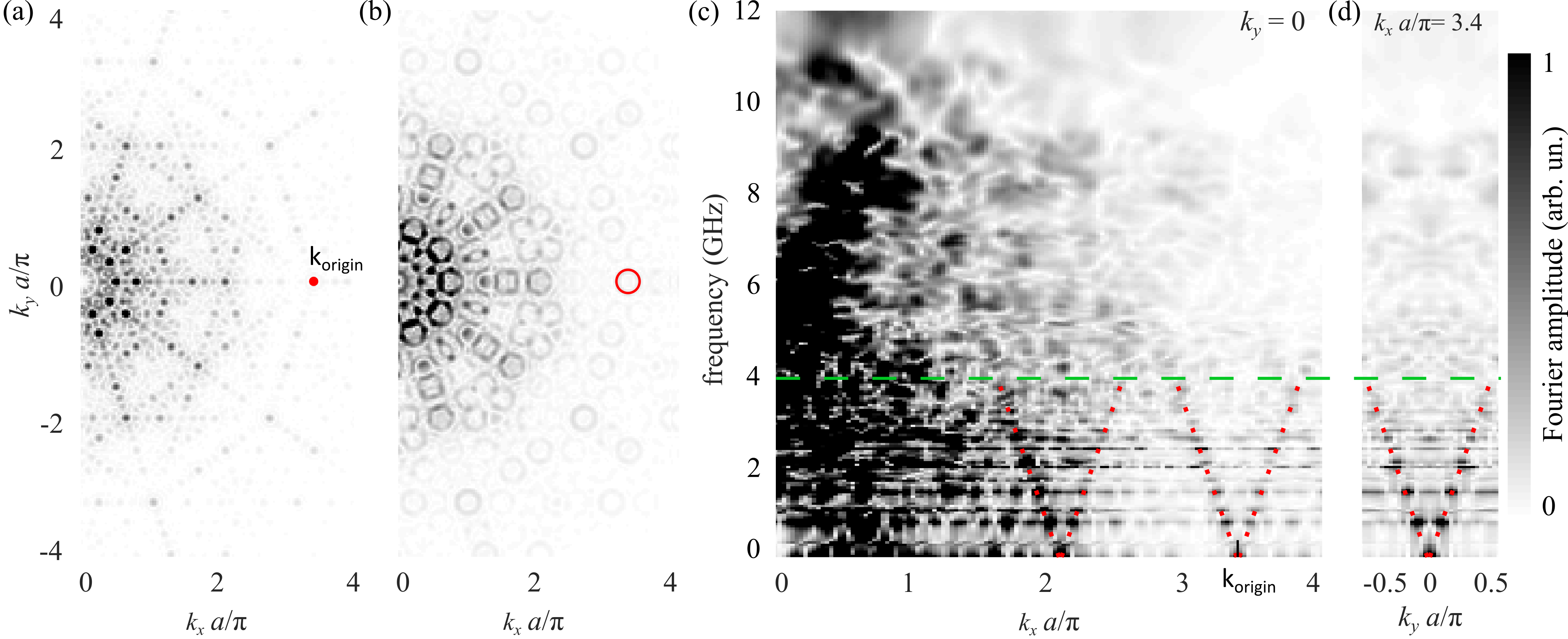}
\caption{\label{fig:Fourier} 
Electric field $E_z$ distribution over the quasicrystalline IWM in the reciprocal (Fourier) space. Reciprocal space at (a) the quasistatic (at 0.1 GHz) and (b) propagation (at 1.5 GHz) regimes. Cross-section of the reciprocal space and frequency axes containing a node at $\mathbf{k}=(3.4,0)\pi/a$ along $\mathbf {k}_x$ axis (c) and $\mathbf {k}_y$ axis (d). The node is marked by red dot in panel (a). The red dotted lines in (c) and (d) highlight the linear isotropic dispersion dependence in the interval below 4 GHz.
}
\end{figure*}

Two-dimensional periodic structures are known to be associated with one of five possible Bravais lattices and identical elements forming the unit cell are located in each site of the lattice. As to the case of quasicrystalline structures in particular Penrose tiling, there exists a kind of generalized Bravais lattice, the nodes of which are yet to be assigned to a certain material structure.  For example, a quasicrystalline metasurface can be based on dielectric disks located in the apex (nodes) of Penrose tiling \cite{kruk2015polarization,neve2010resonant} 
However, quasicrystalline  IWM requires a more complex design, as it should include conductive metallic grids with interconnections between the adjacent cells.
The design of the unit cells implies a pair of non-connected subgrids forming the IWM. Two unit cells of the IWM represented by parallelepipeds (with Penrose tiles as bases) contain the pair of $\Pi$- and U-shaped wires oriented along the diagonals (shown in gray ($\Pi$) and yellow (U) in Fig.~\ref{fig:geometry}b), the space between grids is filled with a host material which in our case is air.
When constructing stars from tiles (one of the seven possible stars is shown in the Fig. \ref{fig:geometry}c), wires in adjoined tiles are connected to each other and form
two electrically independent grids: the first one includes diagonal wires connected to the central node (shown in red), the second one includes diagonal wires surrounding the central node (shown in blue). On the other hand, each of the grids turns out to be in the upper plane of the quasicrystal, then in the lower one. In the static limit, the subgrids have independent electric potentials of $\phi_1$ and $\phi_2$. As a result, the electric field vector inside each tile is directed either up or down and predominantly undergoes an overturn between neighboring tiles. We note that for periodic IWM \cite{chen2018metamaterials} the overturn of E-field takes place with respect to every neighboring site. For the quasicrystalline IWM there are some exceptions. For example, Fig.~\ref{fig:geometry}c shows the adjoining tiles and four of them have identical configuration (red wire in the upper plane and the blue in the bottom plane), but one has inverse configuration (red wire in the bottom plane and the blue in the upper plane). As a result, the electric field in the static limit flips with respect  to neighboring tiles with opposite configuration, as in the case of periodic IWM.

However, the above features do not annihilate the hallmark property of IWM, i.e., the shift of light cones from the origin \cite{chen2018metamaterials, sakhno2021longitudinal} (see below).

Figure~\ref{fig:geometry}a illustrates the procedure of converting Penrose tiling to the quasicrystalline IWM. As a result, the proposed structure is a metasurface which has two flat boundaries (upper and bottom) formed by the two subgrids of interlaced wire medium. 
The described design of wire configuration in the unit cells of Penrose tiling allows generation of a sample with an arbitrarily large size, and what is essential, the sample is composed of two non-connected subgrids.

\section{Electromagnetic properties of quasicrystal IWM}

We study electromagnetic transport of proposed structure by using full wave simulation with the time domain solver of CST Microwave Studio software. 
The source of energy is a discrete port connected between the subgrids at the center of a 665 tiles sample with the diameter of 276 mm (Fig.~\ref{fig:modes} a). The material of wires is perfect electric conductor (PEC), the wire radius is 1 mm, the tile side $a$ is of $10$~mm, the sample thickness is 5 mm and the host media is air.
We found that the quasicrystalline IWM support TE modes, i.e. the dominant component of electric field oscillates along the normal direction ($z$-axis), whereas the in-plane components of the amplitude $E_x$, $E_y$ averaged over each tile tend to vanish.

Our study reveals three frequency intervals: quasistatic at about 0.1 GHz and less, low (0.38 to 4 GHz) and high (4 to 12 GHz) with different electromagnetic properties (see details below). The properties are studied by means of two complementary approaches. The first one is the examination of electric field distribution in real space. It allows estimating the value of the effective refraction index in low frequency range and showing a qualitative difference between the propagation and localization regimes. Besides, we obtain the localization length of the wave in the quasicrystalline IWM in the high frequency interval. The second approach is the analysis of the reciprocal (Fourier) space that also reveals the difference between the quasicrystalline and periodic IWM

\begin{figure*}
\includegraphics[width=\linewidth]{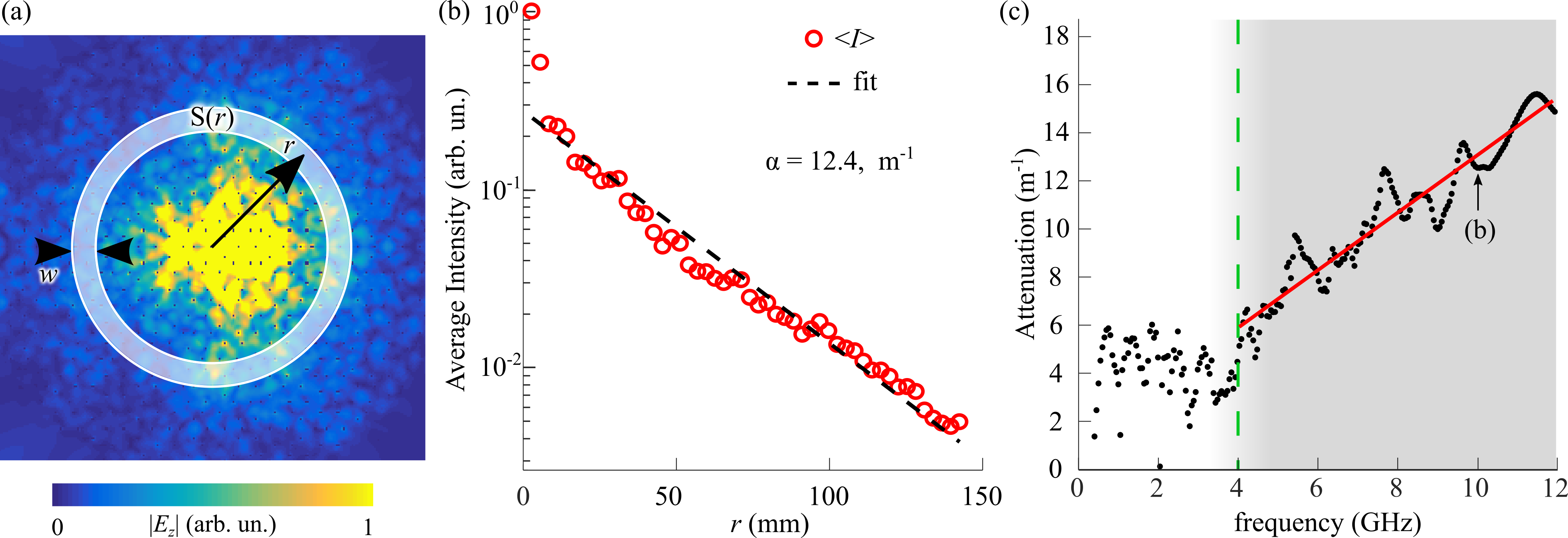}
\caption{\label{fig:Localization} Field pattern in the localization regime. (a) Electric field $E_z$ distribution at 10 GHz. A ring of width $w$ and radius $r$ for the averaging procedure is shaded.
(b) Red circles are angular averaged intensity of the electric field as a function of radius. Black dashed line is the fitting with exponential function $\propto\exp(-\alpha r)$. (c) Spectrum of the attenuation coefficient $\alpha$. The high-frequency interval corresponding to the localization regime is shown by shading.
}
\end{figure*}

\subsection{Analysis in real space}
Figure~\ref{fig:modes}a shows the scheme of our simulations. We study the frequency $f$ dependent power loss factor $P(f)$ of a dipole source located at the center of the sample. The factor is obtained from the simulated S-parameter as $P = 1 - S_{11}^2$, where $S_{11}^2$ is the energy reflected back to the discreet port. The wires are made of PEC so that the coefficient $P$ corresponds to the energy leakage to free space only.

Two well distinguishable frequency ranges are clearly seen in the power loss spectrum (see Fig.~\ref{fig:modes}b). A comb of narrow high-quality (Q) factor peaks is observed in the low-frequency interval (below the frequency of about $f\approx 4$ GHz), whereas the high-frequency interval contains the number of low-Q wide bands. The electric field distributions corresponding to the modes demonstrate the different patterns for the low and high frequency intervals. In the limiting case of the quasistatic regime (Fig.~\ref{fig:modes}c) the field represents the patterns of the Penrose tiling with `rounded’ rhombus unit cells of opposite phases. We observe almost constant amplitude and abrupt phase changing at the rhombus edges due to torsion of subgrids with a spatial displacement from tile to tile.
The patterns of the narrow peaks reveal the recognized Mie-type modes of two-dimensional circular particles (see examples in Figs.~\ref{fig:modes}d and~\ref{fig:modes}e). 
It is important to mention that the field distributions corresponding to the narrow resonance modes are a superposition of rapid oscillation due to IWM design based on Penrose tiling and standing wave pattern due to interference of a propagating wave. The effective refractive index $n$ of IWM is evaluated from the size of the sample (276 mm), the order and the frequency of a particular resonance. Thus for the resonance at 0.38 GHz (Fig. \ref{fig:modes}d) effective refractive index is $n=1.43$, for the mode at 1.5 GHz  (Fig. \ref{fig:modes}e) the effective refractive index is $n=1.45$. Moreover, it should be noted that the refraction index of the quasicrystalline IWM in low-frequency range has no dependence on the direction of propagation, which is seen from the isotropic distribution of the field, 
that is the five-fold rotational symmetry of the Penrose lattice does not lead to anisotropy of the effective refraction index due to the spatial dispersion characteristic of wire media \cite{belov2003strong,simovski2004low}.

\begin{figure}
\includegraphics[width=80 mm]{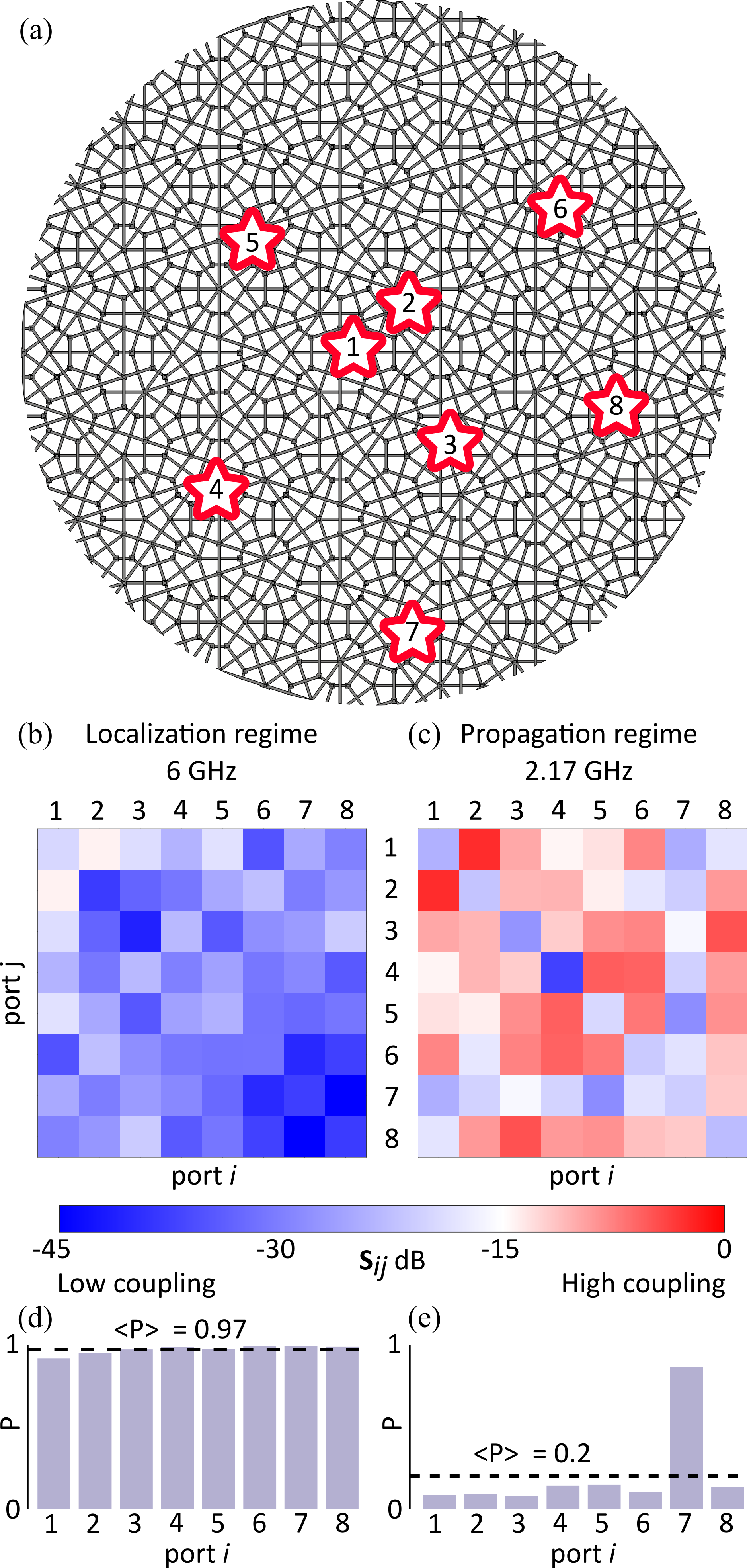}
			\caption{\label{fig:decoupling} 8-channel antenna design. Spatial locations of an arbitrary distributed ports (a). Noise correlation matrices (S-matrices) at 6 GHz (b) and 2.17 GHz (c) include reflection ($i=j$) and coupling ($i\neq j$) coefficients. 
		    Average values of radiation efficiency (power loss) $P_i = 1 - \sum_{j} S_{ij}^2$ were 97$\%$ at 6 GHz (d) and 20$\%$ at 2.17 GHz (e)
			}
\end{figure}

At the high-frequency region, the electric field distribution is poorly structured, neither Penrose tiling nor standing wave patterns are observed. Instead, the electromagnetic energy is localized near the source.
No the wave propagation along the quasicrystalline IWM was found either.
We show below that intensity decay rate of the field increases with the frequency which leads to shrinking of the energy displacement area (compare the distributions shown in Figs.~\ref{fig:modes}f and ~\ref{fig:modes}g).

To determine the type of function and the decay rate, we calculate the average electric field as a function of the distance between the source and the annular area (Fig.~\ref{fig:Localization}a), where the averaging takes place, as follows: $I(r) = \sum_{S(r)}|E_z|/S(r)$, where $S(r)$ is the square of annular region.

As can be seen from Fig.~\ref{fig:Localization}b, the averaged electric field decays exponentially with an attenuation coefficient $\alpha$ as $I(r)\propto e^{-\alpha r}$. Following this procedure, the attenuation coefficient was found for field distributions in the frequency range of 0.1 to 12 GHz with a step of 0.05 GHz (Fig.~\ref{fig:Localization}c). 
In the low-frequency range (below 4 GHz) the root-mean-square deviation from the exponential dependence is significant and the attenuation coefficient undergoes strong fluctuations, so the localization description is not suitable for this range. In the high-frequency range (above 4 GHz), the attenuation coefficient shows almost linear growth with frequency and the deviation from the model becomes weak. Thus, this interval allows the intrinsic wave localization. We note here that such localization is forbidden in structures possessing translation symmetry because of Bloch theorem.

\subsection{Analysis in reciprocal space}

Now we describe the results of the examination of our sample in the reciprocal space, which is another powerful approach in transport properties investigation. According to this technique, the distribution of plane wave amplitudes depending on their wave vectors at a certain frequency is obtained by performing the Fourier transform of the $E_z$ field distribution (Fig.~\ref{fig:Fourier}). Similar to the case of a periodic IWM (\cite{chen2018metamaterials, sakhno2021longitudinal}) its quasicrystalline counterpart studied here supports the spatial harmonics with non-zero wave vectors (see multiple dark spots in Fig.~\ref{fig:Fourier}a) located at the nodes of the quasicrystal lattice in reciprocal space. Also, we notice that the amplitude of Fourier harmonic at the origin $\mathbf{k}=0$ tends to zero in the quasistatic case. 
Ten-fold rotational symmetry $C_{10}$ of the pattern in the reciprocal space in related to the $C_5$ symmetry of our quasicrystalline IWM structure in real space and the time inversion symmetry that has caused the inversion symmetry in the reciprocal space of wave vectors.
Since the $C_{10}$ rotational axis is not compatible with translation symmetry the nodes in the reciprocal space are arranged in a quasicrystal lattice as well. For the quasicrystals, there are no efficient theoretical methods providing us with benefits compared with those following from Bloch theorem, so we just exploit numerical data obtained by Fourier transform.  

It is instructive to analyze the dispersion dependence in the reciprocal space. For structures with translation symmetry each node in the reciprocal space is known to correspond to the apex of the light cone repetition due to uncertainty of the wave vector up to any reciprocal lattice vector. In the present case of an aperiodic structure, we observe the similar behavior of the maximum in the reciprocal space. Let us consider the node at $k_x a/\pi \approx 3.4, k_y = 0$ marked in red in Fig.~\ref{fig:Fourier}a related to almost zero frequency $0.1$~GHz.
Fig.~\ref{fig:Fourier}b shows the reciprocal space at the frequency of 1.5~GHz. The distribution maximum originating at the marked node expanded into a small circle marked in red. Figure ~\ref{fig:Fourier}c (as well as \ref{fig:Fourier}d) shows the cut of reciprocal space along the $k_x$ (and $k_y$) vs. frequency. A light cone typical for isotropic media with apex at the considered node is well defined (see red dotted lines in Figs.~\ref{fig:Fourier}c~and~\ref{fig:Fourier}d). We evaluate the effective refractive index from the slope of the light cone according to the linear dispersion relation (eq. \ref{eq:1})
\begin{equation}
    |\mathbf{k}-\mathbf{k}_\mathrm{node}| = n \frac{\omega}{c}
    \label{eq:1}
\end{equation}
where $\mathbf{k}$ is the wave vector coordinate of the reciprocal space, $\mathbf{k}_\mathrm{node}$ is a node of the quasicrystal lattice in the reciprocal space, $n$ is the effective refractive index, $\omega$ is the angular frequency and $c$ is vacuum speed of light.
The obtained value of the refractive index $n=1.6\pm0.2$ is the same for two cross-sections of the reciprocal space passing through the selected node point.
We notice that it agrees with the values obtained from the Mie mode analysis in real space (see above). Thus, the light properties correspond to the convectional propagation regime of structures allowing homogenization. However, above the frequency of 4 GHz the light cone disappears, which matches the change in the spectrum shown in Fig.~\ref{fig:modes}b.

\section{Multi-channel device}

The localization regime shows two unique properties of quasicrystalline IWM, which are the exponential decay of energy generated by a source and rapid spatial variation of phase. These advantages make it possible to use such a medium in engineering of multi-channel devices, since one of their main requirements is low transmission coefficient between ports located in different positions. It allows to enhance the signal-to-noise ratio and an efficient of beam forming. In particular, such devices are in high demand for applications in 5G communicators \cite{parchin2019eight,li2019high,abdullah2019eight,ojaroudi2019mm} and receiving coils for magnetic resonance imaging \cite{ruytenberg2020shielded,zivkovic2019high,shajan201416,lee2002coupling,yan20147t,connell2014design}. In addition, Micco A. et.al. \cite{micco2009directive} reported that single channel leaky wave antenna made of dielectric rods arranged according to a 12-fold symmetric aperiodic tiling shows broadside radiation at multiple frequencies, with high directivity and low sidelobes. 

We examine a promising multi-channel device based on quasicrystalline IWM. Several ports are connected between the subgrids. The operating frequency range is assumed to be within the high-frequency interval related to the localization regime. We find that the energy radiated from one port (source) reaches the second one (receiver) with a significant attenuation due to both wave localization and radiation to the free space. 

In order to demonstrate transmission between ports in a multi-channel device 8 ports are 
connected between two grids
 at arbitrary locations in the quasicrystalline IWM sample (Fig.~\ref{fig:decoupling}a). We also provided a matching procedure for each port for minimization of the energy reflection back to the transmission line. 
We selected an operation frequency at 6 GHz with the bandwidth of 3$\%$. We notice that the Q-factor in the high-frequency region is very low, which makes it possible to increase bandwidth using broadband matching schemes.
The result is represented as an S-matrix (Fig.~\ref{fig:decoupling}b), which shows both reflection coefficients $S_{ii}$ (signal reflected from the port $i$ back to the transmission line) and coupling coefficients  $S_{ij},\quad i\neq j$ (signal transmitted between ports $i$ and $j$). Obviously, the highest coupling coefficient corresponds to the nearest disposed ports 1 and 2 (distance 26.2 mm or $0.52 \lambda$), but it takes small values and does not exceed -14.2 dB or 3.8$\%$ in energy units.
Such relatively small coupling leads to a high radiation efficiency  (Fig.~\ref{fig:decoupling}d) since energy supplied to a port is radiated predominantly to free space, and just a small portion of it is received by the other ports. In the case of PEC metal, the radiation efficiency is equal to the power loss (Eq. \ref{eq:2}) and the average value over all ports is about 97$\%$.
\begin{equation}
     P_i = 1 - \sum_{j}{S_{ij}^2}\quad \times 100\%
    \label{eq:2}
\end{equation}

We compare this result with that corresponding to the propagation regime which is also available in periodic IWM. The coupling coefficient remains significant and surpasses $-5\, dB$  (Fig.~\ref{fig:decoupling}c) which leads to a dramatic drop down 20$\%$ in radiation efficiency  (Fig.~\ref{fig:decoupling}e). Thus, the localization regime related to the high-frequency interval gives the proposed quasicrystalline IWM to allows unique electromagnetic properties essential for varied applications.

\section{Conclusion}
We have proposed the design of a quasicrystalline IWM structure based on Penrose tiling. The complex analysis of electromagnetic properties by means of two complementary methods in real and reciprocal space has been performed. We have distinguished two frequency intervals with different transport properties. The low-frequency interval corresponds to the propagation regime. The similar transport regime is observed in periodic IWM structures as well. The high-frequency interval relates to the localization regime. Such regime occurs in aperiodic structures only and not possible in structures with translation symmetry. We have discovered that the unique localization regime is beneficial for the design of multi-channel devices appropriate for applications in microwave communication and imaging systems.   

\acknowledgements 

We are thankful to Pavel Belov, Ekaterina Maslova and Artem Sinelnik for fruitful discussions. This work was supported by the Russian Foundation for Basic Research (Grant No. 20-02-00785).


\begin{thebibliography}{37}%
\makeatletter
\providecommand \@ifxundefined [1]{%
 \@ifx{#1\undefined}
}%
\providecommand \@ifnum [1]{%
 \ifnum #1\expandafter \@firstoftwo
 \else \expandafter \@secondoftwo
 \fi
}%
\providecommand \@ifx [1]{%
 \ifx #1\expandafter \@firstoftwo
 \else \expandafter \@secondoftwo
 \fi
}%
\providecommand \natexlab [1]{#1}%
\providecommand \enquote  [1]{``#1''}%
\providecommand \bibnamefont  [1]{#1}%
\providecommand \bibfnamefont [1]{#1}%
\providecommand \citenamefont [1]{#1}%
\providecommand \href@noop [0]{\@secondoftwo}%
\providecommand \href [0]{\begingroup \@sanitize@url \@href}%
\providecommand \@href[1]{\@@startlink{#1}\@@href}%
\providecommand \@@href[1]{\endgroup#1\@@endlink}%
\providecommand \@sanitize@url [0]{\catcode `\\12\catcode `\$12\catcode
  `\&12\catcode `\#12\catcode `\^12\catcode `\_12\catcode `\%12\relax}%
\providecommand \@@startlink[1]{}%
\providecommand \@@endlink[0]{}%
\providecommand \url  [0]{\begingroup\@sanitize@url \@url }%
\providecommand \@url [1]{\endgroup\@href {#1}{\urlprefix }}%
\providecommand \urlprefix  [0]{URL }%
\providecommand \Eprint [0]{\href }%
\providecommand \doibase [0]{http://dx.doi.org/}%
\providecommand \selectlanguage [0]{\@gobble}%
\providecommand \bibinfo  [0]{\@secondoftwo}%
\providecommand \bibfield  [0]{\@secondoftwo}%
\providecommand \translation [1]{[#1]}%
\providecommand \BibitemOpen [0]{}%
\providecommand \bibitemStop [0]{}%
\providecommand \bibitemNoStop [0]{.\EOS\space}%
\providecommand \EOS [0]{\spacefactor3000\relax}%
\providecommand \BibitemShut  [1]{\csname bibitem#1\endcsname}%
\let\auto@bib@innerbib\@empty
\bibitem [{\citenamefont {Shin}\ \emph {et~al.}(2007)\citenamefont {Shin},
  \citenamefont {Shen},\ and\ \citenamefont {Fan}}]{shin2007three}%
  \BibitemOpen
  \bibfield  {author} {\bibinfo {author} {\bibfnamefont {J.}~\bibnamefont
  {Shin}}, \bibinfo {author} {\bibfnamefont {J.-T.}\ \bibnamefont {Shen}}, \
  and\ \bibinfo {author} {\bibfnamefont {S.}~\bibnamefont {Fan}},\ }\href@noop
  {} {\bibfield  {journal} {\bibinfo  {journal} {Physical Review B}\ }\textbf
  {\bibinfo {volume} {76}},\ \bibinfo {pages} {113101} (\bibinfo {year}
  {2007})}\BibitemShut {NoStop}%
\bibitem [{\citenamefont {Latioui}\ and\ \citenamefont
  {Silveirinha}(2017)}]{latioui2017light}%
  \BibitemOpen
  \bibfield  {author} {\bibinfo {author} {\bibfnamefont {H.}~\bibnamefont
  {Latioui}}\ and\ \bibinfo {author} {\bibfnamefont {M.~G.}\ \bibnamefont
  {Silveirinha}},\ }\href@noop {} {\bibfield  {journal} {\bibinfo  {journal}
  {Physical Review B}\ }\textbf {\bibinfo {volume} {96}},\ \bibinfo {pages}
  {195132} (\bibinfo {year} {2017})}\BibitemShut {NoStop}%
\bibitem [{\citenamefont {Sakhno}\ \emph {et~al.}(2021)\citenamefont {Sakhno},
  \citenamefont {Koreshin},\ and\ \citenamefont
  {Belov}}]{sakhno2021longitudinal}%
  \BibitemOpen
  \bibfield  {author} {\bibinfo {author} {\bibfnamefont {D.}~\bibnamefont
  {Sakhno}}, \bibinfo {author} {\bibfnamefont {E.}~\bibnamefont {Koreshin}}, \
  and\ \bibinfo {author} {\bibfnamefont {P.~A.}\ \bibnamefont {Belov}},\
  }\href@noop {} {\bibfield  {journal} {\bibinfo  {journal} {Physical Review
  B}\ }\textbf {\bibinfo {volume} {104}},\ \bibinfo {pages} {L100304} (\bibinfo
  {year} {2021})}\BibitemShut {NoStop}%
\bibitem [{\citenamefont {Chen}\ \emph {et~al.}(2018)\citenamefont {Chen},
  \citenamefont {Hou}, \citenamefont {Zhang}, \citenamefont {Pendry},\ and\
  \citenamefont {Chan}}]{chen2018metamaterials}%
  \BibitemOpen
  \bibfield  {author} {\bibinfo {author} {\bibfnamefont {W.-J.}\ \bibnamefont
  {Chen}}, \bibinfo {author} {\bibfnamefont {B.}~\bibnamefont {Hou}}, \bibinfo
  {author} {\bibfnamefont {Z.-Q.}\ \bibnamefont {Zhang}}, \bibinfo {author}
  {\bibfnamefont {J.~B.}\ \bibnamefont {Pendry}}, \ and\ \bibinfo {author}
  {\bibfnamefont {C.-T.}\ \bibnamefont {Chan}},\ }\href@noop {} {\bibfield
  {journal} {\bibinfo  {journal} {Nature communications}\ }\textbf {\bibinfo
  {volume} {9}},\ \bibinfo {pages} {1} (\bibinfo {year} {2018})}\BibitemShut
  {NoStop}%
\bibitem [{\citenamefont {Shin}\ \emph {et~al.}(2010)\citenamefont {Shin},
  \citenamefont {Shen},\ and\ \citenamefont {Fan}}]{shin2010transmission}%
  \BibitemOpen
  \bibfield  {author} {\bibinfo {author} {\bibfnamefont {J.}~\bibnamefont
  {Shin}}, \bibinfo {author} {\bibfnamefont {J.-T.}\ \bibnamefont {Shen}}, \
  and\ \bibinfo {author} {\bibfnamefont {S.}~\bibnamefont {Fan}},\ }\href@noop
  {} {\bibfield  {journal} {\bibinfo  {journal} {Journal of nanoscience and
  nanotechnology}\ }\textbf {\bibinfo {volume} {10}},\ \bibinfo {pages} {1737}
  (\bibinfo {year} {2010})}\BibitemShut {NoStop}%
\bibitem [{\citenamefont {Powell}\ \emph {et~al.}(2021)\citenamefont {Powell},
  \citenamefont {Mitchell-Thomas}, \citenamefont {Zhang}, \citenamefont
  {Cadman}, \citenamefont {Hibbins},\ and\ \citenamefont
  {Sambles}}]{powell2021dark}%
  \BibitemOpen
  \bibfield  {author} {\bibinfo {author} {\bibfnamefont {A.~W.}\ \bibnamefont
  {Powell}}, \bibinfo {author} {\bibfnamefont {R.~C.}\ \bibnamefont
  {Mitchell-Thomas}}, \bibinfo {author} {\bibfnamefont {S.}~\bibnamefont
  {Zhang}}, \bibinfo {author} {\bibfnamefont {D.~A.}\ \bibnamefont {Cadman}},
  \bibinfo {author} {\bibfnamefont {A.~P.}\ \bibnamefont {Hibbins}}, \ and\
  \bibinfo {author} {\bibfnamefont {J.~R.}\ \bibnamefont {Sambles}},\
  }\href@noop {} {\bibfield  {journal} {\bibinfo  {journal} {ACS photonics}\
  }\textbf {\bibinfo {volume} {8}},\ \bibinfo {pages} {841} (\bibinfo {year}
  {2021})}\BibitemShut {NoStop}%
\bibitem [{\citenamefont {Wang}\ \emph {et~al.}(2021)\citenamefont {Wang},
  \citenamefont {Chen}, \citenamefont {Zetterstrom},\ and\ \citenamefont
  {Quevedo-Teruel}}]{wang2021three}%
  \BibitemOpen
  \bibfield  {author} {\bibinfo {author} {\bibfnamefont {H.}~\bibnamefont
  {Wang}}, \bibinfo {author} {\bibfnamefont {Q.}~\bibnamefont {Chen}}, \bibinfo
  {author} {\bibfnamefont {O.}~\bibnamefont {Zetterstrom}}, \ and\ \bibinfo
  {author} {\bibfnamefont {O.}~\bibnamefont {Quevedo-Teruel}},\ }\href@noop {}
  {\bibfield  {journal} {\bibinfo  {journal} {Applied Sciences}\ }\textbf
  {\bibinfo {volume} {11}},\ \bibinfo {pages} {7153} (\bibinfo {year}
  {2021})}\BibitemShut {NoStop}%
\bibitem [{\citenamefont {Vardeny}\ \emph {et~al.}(2013)\citenamefont
  {Vardeny}, \citenamefont {Nahata},\ and\ \citenamefont
  {Agrawal}}]{vardeny2013optics}%
  \BibitemOpen
  \bibfield  {author} {\bibinfo {author} {\bibfnamefont {Z.~V.}\ \bibnamefont
  {Vardeny}}, \bibinfo {author} {\bibfnamefont {A.}~\bibnamefont {Nahata}}, \
  and\ \bibinfo {author} {\bibfnamefont {A.}~\bibnamefont {Agrawal}},\
  }\href@noop {} {\bibfield  {journal} {\bibinfo  {journal} {Nature Photon.}\
  }\textbf {\bibinfo {volume} {7}},\ \bibinfo {pages} {177} (\bibinfo {year}
  {2013})}\BibitemShut {NoStop}%
\bibitem [{\citenamefont {Dong}\ \emph {et~al.}(2015)\citenamefont {Dong},
  \citenamefont {Chang}, \citenamefont {Huang}, \citenamefont {Hang},
  \citenamefont {Zhong}, \citenamefont {Chen}, \citenamefont {Huang},\ and\
  \citenamefont {Chan}}]{dong2015conical}%
  \BibitemOpen
  \bibfield  {author} {\bibinfo {author} {\bibfnamefont {J.-W.}\ \bibnamefont
  {Dong}}, \bibinfo {author} {\bibfnamefont {M.-L.}\ \bibnamefont {Chang}},
  \bibinfo {author} {\bibfnamefont {X.-Q.}\ \bibnamefont {Huang}}, \bibinfo
  {author} {\bibfnamefont {Z.~H.}\ \bibnamefont {Hang}}, \bibinfo {author}
  {\bibfnamefont {Z.-C.}\ \bibnamefont {Zhong}}, \bibinfo {author}
  {\bibfnamefont {W.-J.}\ \bibnamefont {Chen}}, \bibinfo {author}
  {\bibfnamefont {Z.-Y.}\ \bibnamefont {Huang}}, \ and\ \bibinfo {author}
  {\bibfnamefont {C.~T.}\ \bibnamefont {Chan}},\ }\href@noop {} {\bibfield
  {journal} {\bibinfo  {journal} {Phys. Rev. Lett.}\ }\textbf {\bibinfo
  {volume} {114}},\ \bibinfo {pages} {163901} (\bibinfo {year}
  {2015})}\BibitemShut {NoStop}%
\bibitem [{\citenamefont {Jeon}\ \emph {et~al.}(2017)\citenamefont {Jeon},
  \citenamefont {Kwon},\ and\ \citenamefont {Hur}}]{jeon2017intrinsic}%
  \BibitemOpen
  \bibfield  {author} {\bibinfo {author} {\bibfnamefont {S.-Y.}\ \bibnamefont
  {Jeon}}, \bibinfo {author} {\bibfnamefont {H.}~\bibnamefont {Kwon}}, \ and\
  \bibinfo {author} {\bibfnamefont {K.}~\bibnamefont {Hur}},\ }\href@noop {}
  {\bibfield  {journal} {\bibinfo  {journal} {Nature Phys.}\ }\textbf {\bibinfo
  {volume} {13}},\ \bibinfo {pages} {363} (\bibinfo {year} {2017})}\BibitemShut
  {NoStop}%
\bibitem [{\citenamefont {Sinelnik}\ \emph {et~al.}(2020)\citenamefont
  {Sinelnik}, \citenamefont {Shishkin}, \citenamefont {Yu}, \citenamefont
  {Samusev}, \citenamefont {Belov}, \citenamefont {Limonov}, \citenamefont
  {Ginzburg},\ and\ \citenamefont {Rybin}}]{sinelnik2020experimental}%
  \BibitemOpen
  \bibfield  {author} {\bibinfo {author} {\bibfnamefont {A.~D.}\ \bibnamefont
  {Sinelnik}}, \bibinfo {author} {\bibfnamefont {I.~I.}\ \bibnamefont
  {Shishkin}}, \bibinfo {author} {\bibfnamefont {X.}~\bibnamefont {Yu}},
  \bibinfo {author} {\bibfnamefont {K.~B.}\ \bibnamefont {Samusev}}, \bibinfo
  {author} {\bibfnamefont {P.~A.}\ \bibnamefont {Belov}}, \bibinfo {author}
  {\bibfnamefont {M.~F.}\ \bibnamefont {Limonov}}, \bibinfo {author}
  {\bibfnamefont {P.}~\bibnamefont {Ginzburg}}, \ and\ \bibinfo {author}
  {\bibfnamefont {M.~V.}\ \bibnamefont {Rybin}},\ }\href@noop {} {\bibfield
  {journal} {\bibinfo  {journal} {Adv. Opt. Mater.}\ }\textbf {\bibinfo
  {volume} {8}},\ \bibinfo {pages} {2001170} (\bibinfo {year}
  {2020})}\BibitemShut {NoStop}%
\bibitem [{\citenamefont {Huang}\ and\ \citenamefont
  {Liu}(2018{\natexlab{a}})}]{huang2018quantum}%
  \BibitemOpen
  \bibfield  {author} {\bibinfo {author} {\bibfnamefont {H.}~\bibnamefont
  {Huang}}\ and\ \bibinfo {author} {\bibfnamefont {F.}~\bibnamefont {Liu}},\
  }\href@noop {} {\bibfield  {journal} {\bibinfo  {journal} {Physical review
  letters}\ }\textbf {\bibinfo {volume} {121}},\ \bibinfo {pages} {126401}
  (\bibinfo {year} {2018}{\natexlab{a}})}\BibitemShut {NoStop}%
\bibitem [{\citenamefont {Huang}\ and\ \citenamefont
  {Liu}(2018{\natexlab{b}})}]{huang2018theory}%
  \BibitemOpen
  \bibfield  {author} {\bibinfo {author} {\bibfnamefont {H.}~\bibnamefont
  {Huang}}\ and\ \bibinfo {author} {\bibfnamefont {F.}~\bibnamefont {Liu}},\
  }\href@noop {} {\bibfield  {journal} {\bibinfo  {journal} {Physical Review
  B}\ }\textbf {\bibinfo {volume} {98}},\ \bibinfo {pages} {125130} (\bibinfo
  {year} {2018}{\natexlab{b}})}\BibitemShut {NoStop}%
\bibitem [{\citenamefont {Huang}\ and\ \citenamefont
  {Liu}(2019)}]{huang2019comparison}%
  \BibitemOpen
  \bibfield  {author} {\bibinfo {author} {\bibfnamefont {H.}~\bibnamefont
  {Huang}}\ and\ \bibinfo {author} {\bibfnamefont {F.}~\bibnamefont {Liu}},\
  }\href@noop {} {\bibfield  {journal} {\bibinfo  {journal} {Physical Review
  B}\ }\textbf {\bibinfo {volume} {100}},\ \bibinfo {pages} {085119} (\bibinfo
  {year} {2019})}\BibitemShut {NoStop}%
\bibitem [{\citenamefont {Ara{\'u}jo}\ and\ \citenamefont
  {Andrade}(2019)}]{araujo2019conventional}%
  \BibitemOpen
  \bibfield  {author} {\bibinfo {author} {\bibfnamefont {R.~N.}\ \bibnamefont
  {Ara{\'u}jo}}\ and\ \bibinfo {author} {\bibfnamefont {E.~C.}\ \bibnamefont
  {Andrade}},\ }\href@noop {} {\bibfield  {journal} {\bibinfo  {journal}
  {Physical Review B}\ }\textbf {\bibinfo {volume} {100}},\ \bibinfo {pages}
  {014510} (\bibinfo {year} {2019})}\BibitemShut {NoStop}%
\bibitem [{\citenamefont {Yang}\ \emph {et~al.}(2016)\citenamefont {Yang},
  \citenamefont {Zhang}, \citenamefont {Li}, \citenamefont {Xu}, \citenamefont
  {Singh}, \citenamefont {Liu}, \citenamefont {Li}, \citenamefont {Kruk},
  \citenamefont {Gu}, \citenamefont {Han} \emph {et~al.}}]{yang2016near}%
  \BibitemOpen
  \bibfield  {author} {\bibinfo {author} {\bibfnamefont {Q.}~\bibnamefont
  {Yang}}, \bibinfo {author} {\bibfnamefont {X.}~\bibnamefont {Zhang}},
  \bibinfo {author} {\bibfnamefont {S.}~\bibnamefont {Li}}, \bibinfo {author}
  {\bibfnamefont {Q.}~\bibnamefont {Xu}}, \bibinfo {author} {\bibfnamefont
  {R.}~\bibnamefont {Singh}}, \bibinfo {author} {\bibfnamefont
  {Y.}~\bibnamefont {Liu}}, \bibinfo {author} {\bibfnamefont {Y.}~\bibnamefont
  {Li}}, \bibinfo {author} {\bibfnamefont {S.~S.}\ \bibnamefont {Kruk}},
  \bibinfo {author} {\bibfnamefont {J.}~\bibnamefont {Gu}}, \bibinfo {author}
  {\bibfnamefont {J.}~\bibnamefont {Han}},  \emph {et~al.},\ }\href@noop {}
  {\bibfield  {journal} {\bibinfo  {journal} {Scientific reports}\ }\textbf
  {\bibinfo {volume} {6}},\ \bibinfo {pages} {1} (\bibinfo {year}
  {2016})}\BibitemShut {NoStop}%
\bibitem [{\citenamefont {Chen}\ \emph {et~al.}(2019)\citenamefont {Chen},
  \citenamefont {Xu},\ and\ \citenamefont {Zhou}}]{chen2019topological}%
  \BibitemOpen
  \bibfield  {author} {\bibinfo {author} {\bibfnamefont {R.}~\bibnamefont
  {Chen}}, \bibinfo {author} {\bibfnamefont {D.-H.}\ \bibnamefont {Xu}}, \ and\
  \bibinfo {author} {\bibfnamefont {B.}~\bibnamefont {Zhou}},\ }\href@noop {}
  {\bibfield  {journal} {\bibinfo  {journal} {Physical Review B}\ }\textbf
  {\bibinfo {volume} {100}},\ \bibinfo {pages} {115311} (\bibinfo {year}
  {2019})}\BibitemShut {NoStop}%
\bibitem [{\citenamefont {He}\ \emph {et~al.}(2019)\citenamefont {He},
  \citenamefont {Ding}, \citenamefont {Zhou}, \citenamefont {Wang},\ and\
  \citenamefont {Gong}}]{he2019quasicrystalline}%
  \BibitemOpen
  \bibfield  {author} {\bibinfo {author} {\bibfnamefont {A.-L.}\ \bibnamefont
  {He}}, \bibinfo {author} {\bibfnamefont {L.-R.}\ \bibnamefont {Ding}},
  \bibinfo {author} {\bibfnamefont {Y.}~\bibnamefont {Zhou}}, \bibinfo {author}
  {\bibfnamefont {Y.-F.}\ \bibnamefont {Wang}}, \ and\ \bibinfo {author}
  {\bibfnamefont {C.-D.}\ \bibnamefont {Gong}},\ }\href@noop {} {\bibfield
  {journal} {\bibinfo  {journal} {Physical Review B}\ }\textbf {\bibinfo
  {volume} {100}},\ \bibinfo {pages} {214109} (\bibinfo {year}
  {2019})}\BibitemShut {NoStop}%
\bibitem [{\citenamefont {Duncan}\ \emph {et~al.}(2020)\citenamefont {Duncan},
  \citenamefont {Manna},\ and\ \citenamefont
  {Nielsen}}]{duncan2020topological}%
  \BibitemOpen
  \bibfield  {author} {\bibinfo {author} {\bibfnamefont {C.~W.}\ \bibnamefont
  {Duncan}}, \bibinfo {author} {\bibfnamefont {S.}~\bibnamefont {Manna}}, \
  and\ \bibinfo {author} {\bibfnamefont {A.~E.}\ \bibnamefont {Nielsen}},\
  }\href@noop {} {\bibfield  {journal} {\bibinfo  {journal} {Physical Review
  B}\ }\textbf {\bibinfo {volume} {101}},\ \bibinfo {pages} {115413} (\bibinfo
  {year} {2020})}\BibitemShut {NoStop}%
\bibitem [{\citenamefont {Gr{\"u}nbaum}\ and\ \citenamefont
  {Shephard}(1987)}]{grunbaum1987tilings}%
  \BibitemOpen
  \bibfield  {author} {\bibinfo {author} {\bibfnamefont {B.}~\bibnamefont
  {Gr{\"u}nbaum}}\ and\ \bibinfo {author} {\bibfnamefont {G.~C.}\ \bibnamefont
  {Shephard}},\ }\href@noop {} {\emph {\bibinfo {title} {Tilings and
  patterns}}}\ (\bibinfo  {publisher} {Courier Dover Publications},\ \bibinfo
  {year} {1987})\BibitemShut {NoStop}%
\bibitem [{\citenamefont {Janot}(1994)}]{janot1996quasicrystals}%
  \BibitemOpen
  \bibfield  {author} {\bibinfo {author} {\bibfnamefont {C.}~\bibnamefont
  {Janot}},\ }\href@noop {} {\emph {\bibinfo {title} {Quasicrystals. A
  primer}}}\ (\bibinfo  {publisher} {Clarendon press, Oxford},\ \bibinfo {year}
  {1994})\BibitemShut {NoStop}%
\bibitem [{\citenamefont {Senechal}(1996)}]{senechal1996quasicrystals}%
  \BibitemOpen
  \bibfield  {author} {\bibinfo {author} {\bibfnamefont {M.}~\bibnamefont
  {Senechal}},\ }\href@noop {} {\emph {\bibinfo {title} {Quasicrystals and
  geometry}}}\ (\bibinfo  {publisher} {CUP Archive},\ \bibinfo {year}
  {1996})\BibitemShut {NoStop}%
\bibitem [{\citenamefont {Kruk}\ \emph {et~al.}(2015)\citenamefont {Kruk},
  \citenamefont {Poddubny}, \citenamefont {Powell}, \citenamefont {Helgert},
  \citenamefont {Decker}, \citenamefont {Pertsch}, \citenamefont {Neshev},\
  and\ \citenamefont {Kivshar}}]{kruk2015polarization}%
  \BibitemOpen
  \bibfield  {author} {\bibinfo {author} {\bibfnamefont {S.~S.}\ \bibnamefont
  {Kruk}}, \bibinfo {author} {\bibfnamefont {A.~N.}\ \bibnamefont {Poddubny}},
  \bibinfo {author} {\bibfnamefont {D.~A.}\ \bibnamefont {Powell}}, \bibinfo
  {author} {\bibfnamefont {C.}~\bibnamefont {Helgert}}, \bibinfo {author}
  {\bibfnamefont {M.}~\bibnamefont {Decker}}, \bibinfo {author} {\bibfnamefont
  {T.}~\bibnamefont {Pertsch}}, \bibinfo {author} {\bibfnamefont {D.~N.}\
  \bibnamefont {Neshev}}, \ and\ \bibinfo {author} {\bibfnamefont {Y.~S.}\
  \bibnamefont {Kivshar}},\ }\href@noop {} {\bibfield  {journal} {\bibinfo
  {journal} {Physical Review B}\ }\textbf {\bibinfo {volume} {91}},\ \bibinfo
  {pages} {195401} (\bibinfo {year} {2015})}\BibitemShut {NoStop}%
\bibitem [{\citenamefont {Neve-Oz}\ \emph {et~al.}(2010)\citenamefont
  {Neve-Oz}, \citenamefont {Pollok}, \citenamefont {Burger}, \citenamefont
  {Golosovsky},\ and\ \citenamefont {Davidov}}]{neve2010resonant}%
  \BibitemOpen
  \bibfield  {author} {\bibinfo {author} {\bibfnamefont {Y.}~\bibnamefont
  {Neve-Oz}}, \bibinfo {author} {\bibfnamefont {T.}~\bibnamefont {Pollok}},
  \bibinfo {author} {\bibfnamefont {S.}~\bibnamefont {Burger}}, \bibinfo
  {author} {\bibfnamefont {M.}~\bibnamefont {Golosovsky}}, \ and\ \bibinfo
  {author} {\bibfnamefont {D.}~\bibnamefont {Davidov}},\ }\href@noop {}
  {\bibfield  {journal} {\bibinfo  {journal} {Journal of Applied Physics}\
  }\textbf {\bibinfo {volume} {107}},\ \bibinfo {pages} {063105} (\bibinfo
  {year} {2010})}\BibitemShut {NoStop}%
\bibitem [{\citenamefont {Belov}\ \emph {et~al.}(2003)\citenamefont {Belov},
  \citenamefont {Marques}, \citenamefont {Maslovski}, \citenamefont {Nefedov},
  \citenamefont {Silveirinha}, \citenamefont {Simovski},\ and\ \citenamefont
  {Tretyakov}}]{belov2003strong}%
  \BibitemOpen
  \bibfield  {author} {\bibinfo {author} {\bibfnamefont {P.~A.}\ \bibnamefont
  {Belov}}, \bibinfo {author} {\bibfnamefont {R.}~\bibnamefont {Marques}},
  \bibinfo {author} {\bibfnamefont {S.~I.}\ \bibnamefont {Maslovski}}, \bibinfo
  {author} {\bibfnamefont {I.~S.}\ \bibnamefont {Nefedov}}, \bibinfo {author}
  {\bibfnamefont {M.}~\bibnamefont {Silveirinha}}, \bibinfo {author}
  {\bibfnamefont {C.~R.}\ \bibnamefont {Simovski}}, \ and\ \bibinfo {author}
  {\bibfnamefont {S.~A.}\ \bibnamefont {Tretyakov}},\ }\href@noop {} {\bibfield
   {journal} {\bibinfo  {journal} {Physical Review B}\ }\textbf {\bibinfo
  {volume} {67}},\ \bibinfo {pages} {113103} (\bibinfo {year}
  {2003})}\BibitemShut {NoStop}%
\bibitem [{\citenamefont {Simovski}\ and\ \citenamefont
  {Belov}(2004)}]{simovski2004low}%
  \BibitemOpen
  \bibfield  {author} {\bibinfo {author} {\bibfnamefont {C.}~\bibnamefont
  {Simovski}}\ and\ \bibinfo {author} {\bibfnamefont {P.}~\bibnamefont
  {Belov}},\ }\href@noop {} {\bibfield  {journal} {\bibinfo  {journal}
  {Physical Review E}\ }\textbf {\bibinfo {volume} {70}},\ \bibinfo {pages}
  {046616} (\bibinfo {year} {2004})}\BibitemShut {NoStop}%
\bibitem [{\citenamefont {Parchin}\ \emph {et~al.}(2019)\citenamefont
  {Parchin}, \citenamefont {Al-Yasir}, \citenamefont {Ali}, \citenamefont
  {Elfergani}, \citenamefont {Noras}, \citenamefont {Rodriguez},\ and\
  \citenamefont {Abd-Alhameed}}]{parchin2019eight}%
  \BibitemOpen
  \bibfield  {author} {\bibinfo {author} {\bibfnamefont {N.~O.}\ \bibnamefont
  {Parchin}}, \bibinfo {author} {\bibfnamefont {Y.~I.~A.}\ \bibnamefont
  {Al-Yasir}}, \bibinfo {author} {\bibfnamefont {A.~H.}\ \bibnamefont {Ali}},
  \bibinfo {author} {\bibfnamefont {I.}~\bibnamefont {Elfergani}}, \bibinfo
  {author} {\bibfnamefont {J.~M.}\ \bibnamefont {Noras}}, \bibinfo {author}
  {\bibfnamefont {J.}~\bibnamefont {Rodriguez}}, \ and\ \bibinfo {author}
  {\bibfnamefont {R.~A.}\ \bibnamefont {Abd-Alhameed}},\ }\href@noop {}
  {\bibfield  {journal} {\bibinfo  {journal} {IEEE access}\ }\textbf {\bibinfo
  {volume} {7}},\ \bibinfo {pages} {15612} (\bibinfo {year}
  {2019})}\BibitemShut {NoStop}%
\bibitem [{\citenamefont {Li}\ \emph {et~al.}(2019)\citenamefont {Li},
  \citenamefont {Luo}, \citenamefont {Yang} \emph {et~al.}}]{li2019high}%
  \BibitemOpen
  \bibfield  {author} {\bibinfo {author} {\bibfnamefont {Y.}~\bibnamefont
  {Li}}, \bibinfo {author} {\bibfnamefont {Y.}~\bibnamefont {Luo}}, \bibinfo
  {author} {\bibfnamefont {G.}~\bibnamefont {Yang}},  \emph {et~al.},\
  }\href@noop {} {\bibfield  {journal} {\bibinfo  {journal} {IEEE Transactions
  on Antennas and Propagation}\ }\textbf {\bibinfo {volume} {67}},\ \bibinfo
  {pages} {3820} (\bibinfo {year} {2019})}\BibitemShut {NoStop}%
\bibitem [{\citenamefont {Abdullah}\ \emph {et~al.}(2019)\citenamefont
  {Abdullah}, \citenamefont {Kiani},\ and\ \citenamefont
  {Iqbal}}]{abdullah2019eight}%
  \BibitemOpen
  \bibfield  {author} {\bibinfo {author} {\bibfnamefont {M.}~\bibnamefont
  {Abdullah}}, \bibinfo {author} {\bibfnamefont {S.~H.}\ \bibnamefont {Kiani}},
  \ and\ \bibinfo {author} {\bibfnamefont {A.}~\bibnamefont {Iqbal}},\
  }\href@noop {} {\bibfield  {journal} {\bibinfo  {journal} {IEEE Access}\
  }\textbf {\bibinfo {volume} {7}},\ \bibinfo {pages} {134488} (\bibinfo {year}
  {2019})}\BibitemShut {NoStop}%
\bibitem [{\citenamefont {Ojaroudi~Parchin}\ \emph {et~al.}(2019)\citenamefont
  {Ojaroudi~Parchin}, \citenamefont {Alibakhshikenari}, \citenamefont
  {Jahanbakhsh~Basherlou}, \citenamefont {A~Abd-Alhameed}, \citenamefont
  {Rodriguez},\ and\ \citenamefont {Limiti}}]{ojaroudi2019mm}%
  \BibitemOpen
  \bibfield  {author} {\bibinfo {author} {\bibfnamefont {N.}~\bibnamefont
  {Ojaroudi~Parchin}}, \bibinfo {author} {\bibfnamefont {M.}~\bibnamefont
  {Alibakhshikenari}}, \bibinfo {author} {\bibfnamefont {H.}~\bibnamefont
  {Jahanbakhsh~Basherlou}}, \bibinfo {author} {\bibfnamefont {R.}~\bibnamefont
  {A~Abd-Alhameed}}, \bibinfo {author} {\bibfnamefont {J.}~\bibnamefont
  {Rodriguez}}, \ and\ \bibinfo {author} {\bibfnamefont {E.}~\bibnamefont
  {Limiti}},\ }\href@noop {} {\bibfield  {journal} {\bibinfo  {journal}
  {Applied Sciences}\ }\textbf {\bibinfo {volume} {9}},\ \bibinfo {pages} {978}
  (\bibinfo {year} {2019})}\BibitemShut {NoStop}%
\bibitem [{\citenamefont {Ruytenberg}\ \emph {et~al.}(2020)\citenamefont
  {Ruytenberg}, \citenamefont {Webb},\ and\ \citenamefont
  {Zivkovic}}]{ruytenberg2020shielded}%
  \BibitemOpen
  \bibfield  {author} {\bibinfo {author} {\bibfnamefont {T.}~\bibnamefont
  {Ruytenberg}}, \bibinfo {author} {\bibfnamefont {A.}~\bibnamefont {Webb}}, \
  and\ \bibinfo {author} {\bibfnamefont {I.}~\bibnamefont {Zivkovic}},\
  }\href@noop {} {\bibfield  {journal} {\bibinfo  {journal} {Magnetic resonance
  in medicine}\ }\textbf {\bibinfo {volume} {83}},\ \bibinfo {pages} {1135}
  (\bibinfo {year} {2020})}\BibitemShut {NoStop}%
\bibitem [{\citenamefont {Zivkovic}\ \emph {et~al.}(2019)\citenamefont
  {Zivkovic}, \citenamefont {Teeuwisse}, \citenamefont {Slobozhanyuk},
  \citenamefont {Nenasheva},\ and\ \citenamefont {Webb}}]{zivkovic2019high}%
  \BibitemOpen
  \bibfield  {author} {\bibinfo {author} {\bibfnamefont {I.}~\bibnamefont
  {Zivkovic}}, \bibinfo {author} {\bibfnamefont {W.}~\bibnamefont {Teeuwisse}},
  \bibinfo {author} {\bibfnamefont {A.}~\bibnamefont {Slobozhanyuk}}, \bibinfo
  {author} {\bibfnamefont {E.}~\bibnamefont {Nenasheva}}, \ and\ \bibinfo
  {author} {\bibfnamefont {A.}~\bibnamefont {Webb}},\ }\href@noop {} {\bibfield
   {journal} {\bibinfo  {journal} {Journal of Magnetic Resonance}\ }\textbf
  {\bibinfo {volume} {299}},\ \bibinfo {pages} {59} (\bibinfo {year}
  {2019})}\BibitemShut {NoStop}%
\bibitem [{\citenamefont {Shajan}\ \emph {et~al.}(2014)\citenamefont {Shajan},
  \citenamefont {Kozlov}, \citenamefont {Hoffmann}, \citenamefont {Turner},
  \citenamefont {Scheffler},\ and\ \citenamefont {Pohmann}}]{shajan201416}%
  \BibitemOpen
  \bibfield  {author} {\bibinfo {author} {\bibfnamefont {G.}~\bibnamefont
  {Shajan}}, \bibinfo {author} {\bibfnamefont {M.}~\bibnamefont {Kozlov}},
  \bibinfo {author} {\bibfnamefont {J.}~\bibnamefont {Hoffmann}}, \bibinfo
  {author} {\bibfnamefont {R.}~\bibnamefont {Turner}}, \bibinfo {author}
  {\bibfnamefont {K.}~\bibnamefont {Scheffler}}, \ and\ \bibinfo {author}
  {\bibfnamefont {R.}~\bibnamefont {Pohmann}},\ }\href@noop {} {\bibfield
  {journal} {\bibinfo  {journal} {Magnetic resonance in medicine}\ }\textbf
  {\bibinfo {volume} {71}},\ \bibinfo {pages} {870} (\bibinfo {year}
  {2014})}\BibitemShut {NoStop}%
\bibitem [{\citenamefont {Lee}\ \emph {et~al.}(2002)\citenamefont {Lee},
  \citenamefont {Giaquinto},\ and\ \citenamefont {Hardy}}]{lee2002coupling}%
  \BibitemOpen
  \bibfield  {author} {\bibinfo {author} {\bibfnamefont {R.~F.}\ \bibnamefont
  {Lee}}, \bibinfo {author} {\bibfnamefont {R.~O.}\ \bibnamefont {Giaquinto}},
  \ and\ \bibinfo {author} {\bibfnamefont {C.~J.}\ \bibnamefont {Hardy}},\
  }\href@noop {} {\bibfield  {journal} {\bibinfo  {journal} {Magnetic Resonance
  in Medicine: An Official Journal of the International Society for Magnetic
  Resonance in Medicine}\ }\textbf {\bibinfo {volume} {48}},\ \bibinfo {pages}
  {203} (\bibinfo {year} {2002})}\BibitemShut {NoStop}%
\bibitem [{\citenamefont {Yan}\ \emph {et~al.}(2014)\citenamefont {Yan},
  \citenamefont {Zhang}, \citenamefont {Feng}, \citenamefont {Ma},
  \citenamefont {Wei},\ and\ \citenamefont {Xue}}]{yan20147t}%
  \BibitemOpen
  \bibfield  {author} {\bibinfo {author} {\bibfnamefont {X.}~\bibnamefont
  {Yan}}, \bibinfo {author} {\bibfnamefont {X.}~\bibnamefont {Zhang}}, \bibinfo
  {author} {\bibfnamefont {B.}~\bibnamefont {Feng}}, \bibinfo {author}
  {\bibfnamefont {C.}~\bibnamefont {Ma}}, \bibinfo {author} {\bibfnamefont
  {L.}~\bibnamefont {Wei}}, \ and\ \bibinfo {author} {\bibfnamefont
  {R.}~\bibnamefont {Xue}},\ }\href@noop {} {\bibfield  {journal} {\bibinfo
  {journal} {IEEE transactions on medical imaging}\ }\textbf {\bibinfo {volume}
  {33}},\ \bibinfo {pages} {1781} (\bibinfo {year} {2014})}\BibitemShut
  {NoStop}%
\bibitem [{\citenamefont {Connell}\ \emph {et~al.}(2014)\citenamefont
  {Connell}, \citenamefont {Gilbert}, \citenamefont {Abou-Khousa},\ and\
  \citenamefont {Menon}}]{connell2014design}%
  \BibitemOpen
  \bibfield  {author} {\bibinfo {author} {\bibfnamefont {I.~R.}\ \bibnamefont
  {Connell}}, \bibinfo {author} {\bibfnamefont {K.~M.}\ \bibnamefont
  {Gilbert}}, \bibinfo {author} {\bibfnamefont {M.~A.}\ \bibnamefont
  {Abou-Khousa}}, \ and\ \bibinfo {author} {\bibfnamefont {R.~S.}\ \bibnamefont
  {Menon}},\ }\href@noop {} {\bibfield  {journal} {\bibinfo  {journal} {IEEE
  transactions on medical imaging}\ }\textbf {\bibinfo {volume} {34}},\
  \bibinfo {pages} {836} (\bibinfo {year} {2014})}\BibitemShut {NoStop}%
\bibitem [{\citenamefont {Micco}\ \emph {et~al.}(2009)\citenamefont {Micco},
  \citenamefont {Galdi}, \citenamefont {Capolino}, \citenamefont {Della~Villa},
  \citenamefont {Pierro}, \citenamefont {Enoch},\ and\ \citenamefont
  {Tayeb}}]{micco2009directive}%
  \BibitemOpen
  \bibfield  {author} {\bibinfo {author} {\bibfnamefont {A.}~\bibnamefont
  {Micco}}, \bibinfo {author} {\bibfnamefont {V.}~\bibnamefont {Galdi}},
  \bibinfo {author} {\bibfnamefont {F.}~\bibnamefont {Capolino}}, \bibinfo
  {author} {\bibfnamefont {A.}~\bibnamefont {Della~Villa}}, \bibinfo {author}
  {\bibfnamefont {V.}~\bibnamefont {Pierro}}, \bibinfo {author} {\bibfnamefont
  {S.}~\bibnamefont {Enoch}}, \ and\ \bibinfo {author} {\bibfnamefont
  {G.}~\bibnamefont {Tayeb}},\ }\href@noop {} {\bibfield  {journal} {\bibinfo
  {journal} {Physical Review B}\ }\textbf {\bibinfo {volume} {79}},\ \bibinfo
  {pages} {075110} (\bibinfo {year} {2009})}\BibitemShut {NoStop}%
\end{thebibliography}

%
\end{document}